\documentclass[12pt,preprint]{aastex}

\shorttitle{X-ray signature from coronal electron beams}
\shortauthors{Saint-Hilaire et al.}

\begin{document}

\title{On the X-ray detectability of electron beams escaping from the Sun}

\author{Pascal Saint-Hilaire\altaffilmark{1}, S\"am Krucker\altaffilmark{1}, Steven Christe\altaffilmark{1}, and Robert P. Lin\altaffilmark{1}}
\affil{Space Sciences Laboratory, University of California,
    Berkeley, CA 94720}

\email{shilaire@ssl.berkeley.edu}

\begin{abstract}
	
	We study the detectability and characterisation of electron beams as they leave their acceleration site in the low corona towards interplanetary space,
	through their non-thermal X-ray bremsstrahlung emission. 
	We demonstrate that the largest interplanetary electron beams ($\gtrsim$10$^{35}$ electrons above 10 keV) can be detected in X-rays with current and future instrumentation,
	such as RHESSI or XRT onboard Hinode.
	We make a list of optimal observing conditions and beam characteristics.
	Amongst others, good imaging (as opposed to mere localisation or detection in spatially-integrated data) is required for proper characterization, 
	putting the requirement on the number of escaping electrons (above 10 keV) to $\gtrsim$3$\times$10$^{36}$ for RHESSI;
	$\gtrsim$3$\times$10$^{35}$ for Hinode/XRT; 
	and $\gtrsim$10$^{33}$ electrons for the FOXSI sounding rocket scheduled to fly in 2011.
	Moreover, we have found that simple modeling hints at the possibility that coronal soft X-ray jets could be the result of local heating by propagating electron beams.
	
\end{abstract}



\section{Introduction}

	In the standard flare scenario, an acceleration site in the corona (a few Mm to a few tens of Mm above the photosphere), 
	generates energetic electrons which propagate along magnetic field lines, either towards the lower corona/chromosphere, or into interplanetary space.
	As they propagate, they lose energy via Coulomb collisions, and perhaps also via wave-particle interaction.
	Energy losses by bremsstrahlung or magnetobremsstrahlung are negligible at our energies of interest ($\approx$1--100 keV).

	During close encounters with ambiant ions, electrons emit hard X-rays (HXR) by bremsstrahlung.
	There are numerous observations of HXR emission in the solar corona during flares \citep[see e.g.][]{Dennis1985}, 
	but so far none of them have been successfully associated with beams of electrons propagating outwards in the tenuous corona \citep[e.g.][]{Christe2008}.
	Unless particle trapping occurs \citep[as is thought to occur in coronal sources, see][and references therein]{Krucker2008}, 
	such observations are indeed difficult to make, given the small column densities electron beams encounter in the corona.

	Two sets of independent observations support the existence of outward-going coronal electron beams: 
	(a) Type III radio bursts which occur simultaneously with HXR emission during the impulsive phase of the flare, and whose frequency decreases with time:
	these are interpreted as radiation caused by electron beams which excite plasma emission in the increasingly tenuous coronal plasma as they travel outwards \citep[see e.g.][]{Dulk1985,Bastian1998}.
	(b) Interplanetary electrons are detected {\it in situ} at 1 AU \citep[e.g.][]{Lin1985}.
	Their onset can often be traced back to a flaring time, when their acceleration is thought to occur \citep{Lin1971,Krucker2007}.

	Reconnection theory \citep[ideal MHD, e.g.][and references therein]{Priest2002} generally predicts that magnetic reconnection is symmetrical about the X-point:
	Assuming that such reconnection is the mechanism responsible for particle acceleration, then it seems reasonable to expect that
	the downward-going beam (that will stop in the chromosphere) and the upward-going beam (that will later escape into interplanetary space) have similar characteristics.
	Surprisingly, the number of escaping interplanetary electrons seem to be only about 0.1\%--1\% that of the X-ray producing electrons precipitating in the chromosphere \citep[see e.g.][]{Lin1971,Krucker2007}.
	Estimates on the number of electrons required to produce a radio type III burst are difficult to obtain.
	\citet{Wentzel1982} used a value of 10$^{33}$ electrons in his discussion of possible theories of Type III bursts.
	\citet{Lin1971} have reported 10$^{33}$--10$^{34}$ electrons above 22 keV for interplanetary beams (in situ observations), 
	and \citet{Kane1972} showed from the upper limit on the flux of thin-target X-rays that less than 10$^{34}$ above 22 keV were required to produce a strong Type III burst at 500 MHz.
	There is also evidence \citep[see e.g.][]{Benz1982,Dennis1984} that there exist a secondary acceleration site for Type III-producing electrons, 
	somehow triggered by the primary energy release: for example, through narrow-band electromagnetic waves from the precipitating flare electrons \citep{Sprangle1983},
	or through another, secondary, reconnection process high in the corona \citep{Vrsnak2003}.
	
	The goal of this paper is to numerically estimate the amount (spatial and spectral distribution) of HXRs emitted by electron beams as they propagate,
	and determine the ability of various space-borne instruments, namely: RHESSI \citep[Ramaty High Energy Solar Spectroscopic Imager,][]{Lin2002}, GOES, {\it Hinode}/XRT \citep{Golub2007}, 
	and the upcoming FOXSI\footnote{The FOXSI (Focusing Optics X-ray Solar Imager) is a recently accepted rocket flight proposal under the ``Low Cost Access to Space'' (LCAS) NASA program, 
	which will use grazing incidence mirrors to focus hard X-rays} rocket flight \citep[which will use HXR focusing optics, see e.g.][]{Ramsey2000}, to observe and identify X-ray emission from such electron beams, 
	at the moment that they exit the presumed acceleration site near the solar surface.
		

\section{Framework}
	The injected (accelerated) beams of electrons are assumed to be power-laws, with a low-energy cutoff $E_1$:
	 
		\begin{eqnarray} \label{eq:F0}
			F_0(E)	& = &	\left\{	\begin{array}{ll} 
						(\delta-1) \frac{F_1}{E_1} \left( \frac{E}{E_1} \right)^{-\delta}	& ,  \,\,\, E > E_1	\\
						0									& , \,\,\, E < E_1	\\
						\end{array} \right.
		\end{eqnarray}
	The distribution $F_0(E)$ is expressed in electrons s$^{-1}$ keV$^{-1}$, 
	$\delta$ is the spectral index, and $F_1$ the total number of injected electrons per second above $E_1$ \citep[same notations as in][]{Brown2002}.

	\citet{PSH2002}, have associated the HXR emission from a flare (GOES class C9.6) with a beam of electrons propagating downward, toward the denser chromosphere,
	and, assuming a thick-target model \citep{Brown1971}, have found the following characteristics for the injected electron beam:
	$\delta$=4, $E_1$=10 keV, and $F_1$=2.7$\times 10^{36}$ electrons/s.
	Despite its relatively small X-ray thermal footprint, this flare was particularly hard, and was even a gamma-ray line emitter.
	For comparison, the 2002 July 23 flare (GOES class X4.8), had an average electron flux of about 10$^{35}$ electrons/s (electrons above 35 keV) during its $\approx$15-minute long main impulsive phase \citep{Holman2003},
	translating to about 10$^{36}$ electrons above 10 keV per second (using an averaged electron spectral index $\delta$ of $\approx$2.5).
	In situ and remote observations from \citet{Krucker2007} indicate that the number of electrons in interplanetary beams is $\approx$0.2\% of the number derived from the temporally associated HXR flare beams.
	This relationship was established for electrons $>$50 keV. 
	Assuming it holds for energies down to 10 keV, this means that the interplanetary counterpart of the first flare beam has $F_1\approx$ 10$^{34}$ electrons/s above 10 keV.
	We will henceforth call a {\it strong} beam a beam of electrons with $F_{1,strong}$=2.7$\times 10^{36}$ electrons/s, and a {\it weak} beam one with $F_{1,weak}$=1.0$\times 10^{34}$ electrons/s ($F_{1,weak} \approx 0.0037 \times  F_{1,strong}$).
	A {\it strong} beam is of the type usually associated with flares, whereas a {\it weak} beam is of the type usually associated with Solar Energetic Particles (SEP).
	The HXR-producing electron beams in \citet{Krucker2007} had a typical duration of 100 s, and this is the duration that will be used throughout this paper, unless otherwise specified.
	For comparison, the 100 s typical duration leads to a total number of {\it weak} beam electrons above 10 keV of $\approx$1$\times$10$^{36}$ electrons, or $\approx$9$\times$10$^{34}$ electrons above 22 keV,
	hence about an order of magnitude more than has been reported so far \citep[e.g.][]{Lin1971,Kane1972}, but still below the 4$\times$10$^{36}$ electrons above 10 keV that would have come out of the 
	2002 July 23 flare, assuming the $\approx$0.2\% relationship holds.

	As in \citet{Brown2002} (and using the same assumptions), the electron spectrum changes shape as it propagates, due to Coulomb energy losses, according to:

		\begin{equation}
			F(E,N)= \frac{E}{\sqrt{E^2 + 2KN}} \,\, F_0 \left( \sqrt{E^2 + 2KN} \right) \\ \label{eq:FEN1}
		\end{equation}
		\begin{equation}
			=\left\{	\begin{array}{ll} 
						(\delta-1) \frac{F_1}{E_1} E_1^{\delta}	\,\, E (E^2 + 2KN)^{-\frac{\delta+1}{2}} 	& \,\,\, ,E > \zeta	\\
						0 											& \,\,\, ,E < \zeta	\\
					\end{array} \right.
		\end{equation}\\ \label{eq:FEN2}

	\noindent where $N$ is the electron column density traversed by the beam of electrons, $K$=2.6$\times 10^{-18}$ cm$^2$ keV$^2$, 
	and $\zeta$=$\sqrt{\max(0, E_1^2-2KN)}$ is the position of the low-energy cutoff after a column density $N$ has been traversed.

	The bremsstrahlung emission per unit column density, along the path of propagation is \citep[from][]{Brown2002}:
		\begin{equation}	\label{eq:int}
			\frac{dI}{dN}(\varepsilon,N) = \frac{1}{4\pi D^2} \int_\varepsilon^\infty F(E,N) \, Q(\varepsilon,E) \, dE
		\end{equation}
	in photons/s/cm$^{-2}$/keV, where $D$ is 1 AU, $N$ is the column density already traversed by the electron beam, 
	and $Q(\varepsilon,E)$ is the differential (for emitted photon energy $\varepsilon$) bremsstrahlung cross-section.

	Using the Kramers cross-section yields an analytical solution to Eq.~(\ref{eq:int}) \citep[][Appendix~\ref{appendix:dIdN}]{Brown2002}, but does not yield an accurate photon spectrum below the low-energy cutoff.
	Hence, numerical evaluations of $\frac{dI}{dN}$ using Eq.~(\ref{eq:int}) and the more proper non-relativistic Bethe-Heitler differential bremsstrahlung cross-section (see appendix \ref{appendix:dIdN}) were used, in order to cover the general case.

	Finally, it must be mentioned that we will exclusively use an isotropic bremsstrahlung cross-section.
	In reality, bremsstrahlung X-ray emission has anisotropic directivity.
	The exact details depend on the spectral shape of the electron energy distribution, the thickness of the target, the pitch angle distribution, the angles between the electron beam trajectory, the magnetic field and the observer, and the energy of the emitted X-rays.
	\citet{Elwert1971} (thin-target case) and \citet{Brown1972} (thick-target case, applicable in our case low emitted photon energies: $\varepsilon\lesssim$10 keV),
	have both concluded a general limb brightening effect \citep[for more recent work on the topic, see also][]{Massone2004}.
	This limb-brightening effect is small at low energies (electron beams at the limb actually produce $\approx$50\% more 10-keV photons than the isotropic bremsstrahlumg cross-section amount), 
	and increases with photon energy ($\approx$100\% more 50-keV thin-target photons than if assuming an isotropic bremsstrahlung cross-section).
	Hence, the effects of bremsstrahlung cross-section anisotropy actually play in our favor, as we are mostly interested in electron beams beyond the solar limb.
	
\section{Numerical simulations}
	
	The ambiant density is the most important factor contributing to HXR fluxes from beams of electrons in the corona, besides the total number of electrons involved.
	It must be of a sufficiently high value.

\begin{figure*}[ht!]
		\centering
		\includegraphics[width=16.6cm]{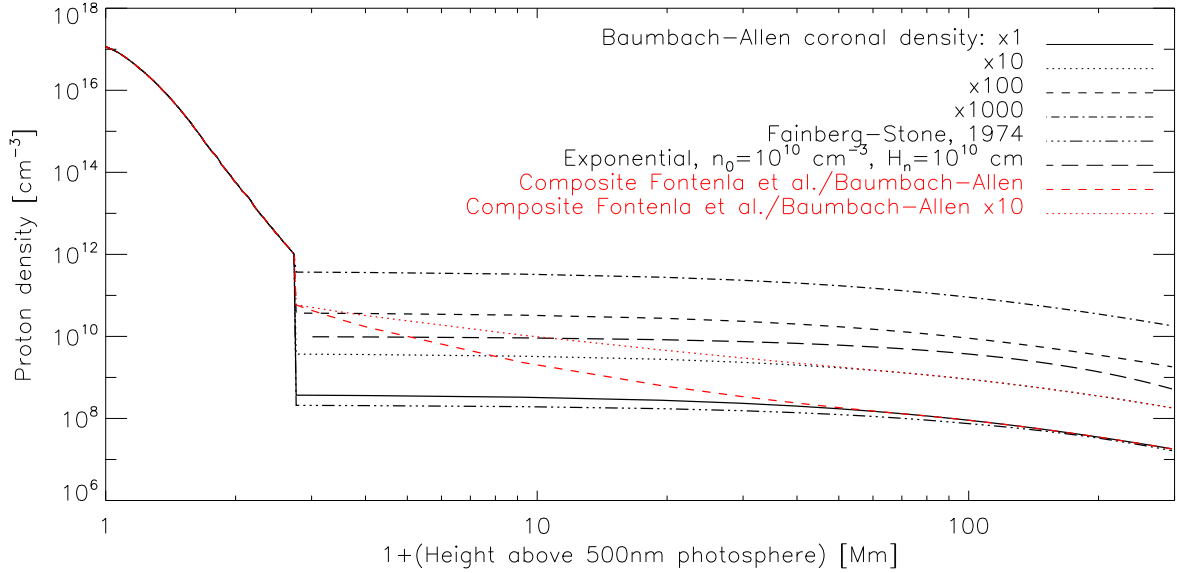}		
		\caption{
			From the transition region and below, densities from \citet{Fontenla1993} (model ``P'') were used.
			The \citet{Fainberg1974} coronal density model is essentially the same as the Baumbach-Allen one, and, when near the solar surface, 
			both are very similar to a barometric atmosphere with density scale height $H_n$=10$^{10}$ cm.
			As coronal streamers can be an order of magnitude denser, coronas with higher densities (x10, x100, x1000) were also considered.
			The {\it red} models are artificial composites, where the last data point in the Fontenla et al.'s ``P'' model
			is connected to the Baumbach-Allen coronal density at 69.6 Mm of altitude (0.1 solar radius).
		}
		\label{fig:density}
		\end{figure*}

	Type III radio bursts, which are believed to be caused by electron beams through the bump-on-tail plasma instability and Langmuir wave conversion into EM waves, 
	often start as high as 500 MHz, corresponding to a plasma density of $\sim 3\times10^9$ cm$^{-3}$.
	Start ambiant densities for coronal/interplanetary electron beams can hence be inferred to be as high as this 
	(and even higher, as the electron beam may propagate some distance beyond its injection site before conditions for plasma emission allow the generation of a Type III radio burst).
	When ten times the standard Baumbach-Allen \citep{Baumbach1937} coronal density structure is assumed (coronal streamers can be an order of magnitude (or more) denser than the quiet corona \citep{Fainberg1974}),
	this density of $\sim 3\times10^9$ cm$^{-3}$ corresponds roughly to an acceleration altitude of 20 Mm (see Fig.~\ref{fig:density}). 
	Unless otherwise specified, these are the numbers used in our calculations.

\subsection{Spectra, profiles, and RHESSI imaging}

\begin{figure*}[ht!]
		\centering
		\includegraphics[width=12.0cm]{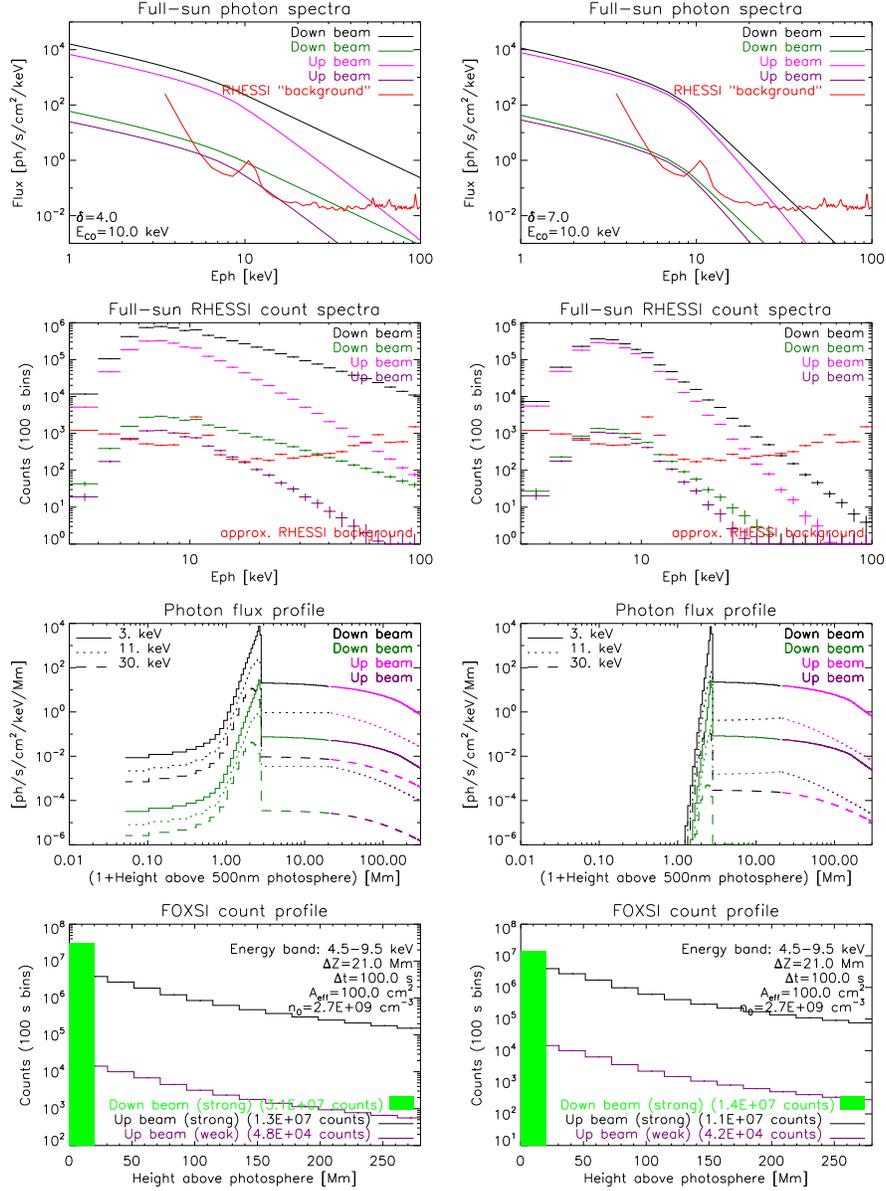}	
		\caption{
			For either {\it strong} (2.7$\times$10$^{36}$ electrons/s) and {\it weak} (1.0$\times$10$^{34}$ electrons/s) beams, 
			going either ``down'' or ``up'', and for $E_1$=10 keV, the following were calculated:
			{\it First row}: Full-Sun photon spectra, including a typical RHESSI ``background'' spectrum {\it (red)}.
			{\it Second row}: Full-Sun RHESSI count spectra, with typical RHESSI background.
			{\it Third row}: Photon flux profiles.
			{\it Fourth row}:  FOXSI count profiles.
			{\it Left column:} $\delta$=4. {\it Right column:} $\delta$=7.
			Statistical errors (barely noticeable on these graphs) have been included for both RHESSI count spectra and FOXSI count profiles,
			for which a 100 s accumulation time (a typical flare duration) and a 5-keV wide band centered around 7 keV were considered.
		}
		\label{fig:sp_prof_Eco10}
		\end{figure*}

	From the injected beam characteristics and the atmospheric density profile, the spatially-integrated photon spectra, 
	RHESSI count spectra, photon flux spatial profiles, 
	and FOXSI countrate profiles, 
	such as in  Fig. \ref{fig:sp_prof_Eco10}\footnote{\small Much more at \url{http://sprg.ssl.berkeley.edu/ $\sim$shilaire/work/ebeam\_June2007/wwwoutput/browser2.html}}, 
	can be calculated (a typical flare duration of 100 s was taken to compute the count rates).
	
	In cases where the ``down'' beam is {\it weak} ({\it green curve} in Fig.~\ref{fig:sp_prof_Eco10}) or nonexistant (such as when flare footpoints are occulted by the solar limb), 
	a {\it strong} ``up'' beam ({\it red curve} in Fig.~\ref{fig:sp_prof_Eco10}) should be easily observed in a RHESSI full-sun spectrum and by FOXSI.
	A {\it weak} ``up'' beam ({\it dark purple curve} in Fig.~\ref{fig:sp_prof_Eco10}) is barely noticeable in a RHESSI full-sun spectra, but still well within FOXSI imaging capabilities.

\begin{figure*}[ht!]
		\centering
		\includegraphics[height=13.8cm]{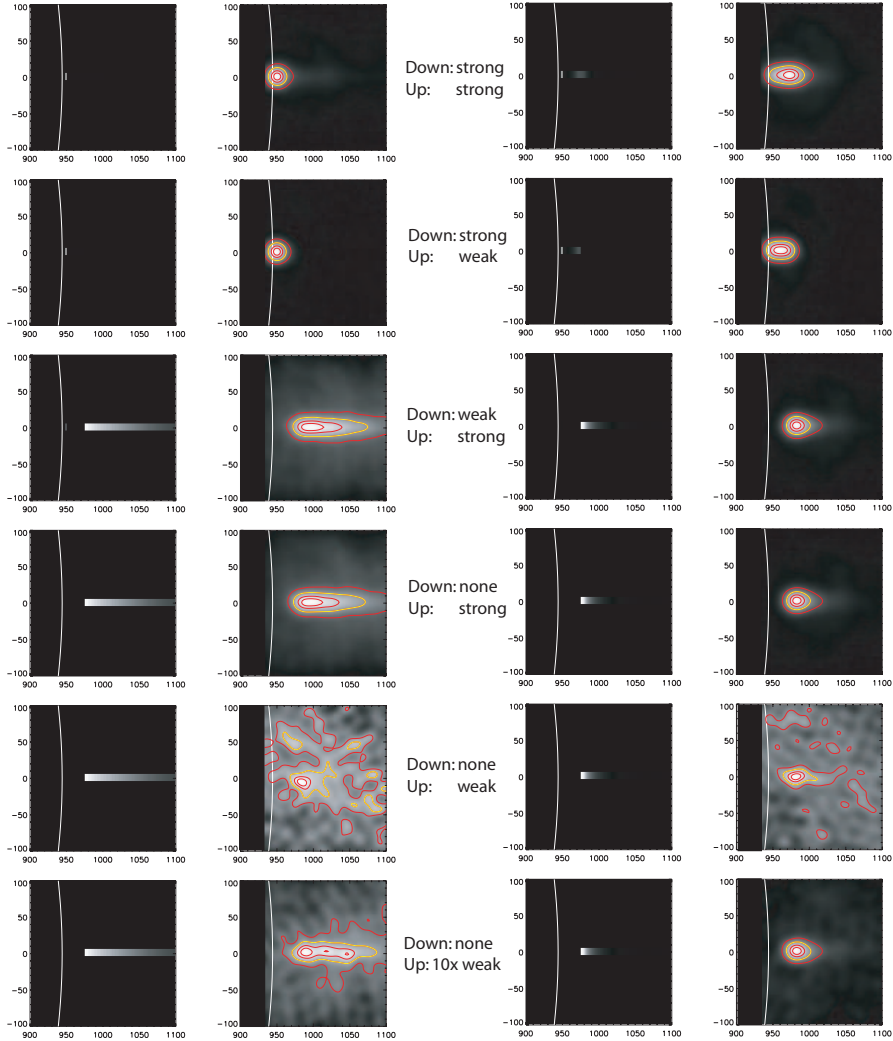}		
		\caption{
			Imaged HXR from electrons beams (1 Mm in diameter), in the 5-9 keV band (optimal RHESSI sensitivity).
			{\it Left column pair:} using 10$\times$ Baumbach-Allen coronal density model (typical in coronal streamers).
			{\it Right column pair:} using 100$\times$ Baumbach-Allen coronal density model.
			{\it Left column of each column pair:} theoretical images.
			{\it Right column of each column pair:} RHESSI simulated image, using detectors 3-9 and the CLEAN algorithm ($\approx$10'' resolution).
			{\it Yellow line}: 50\% contour, {\it red lines}: 25, 75, and 90\% contours.
			Both ``down'' and ``up'' beam start at 20'' altitude above the photosphere ({\it solid white line}),
			corresponding to a density of 3$\times$10$^9$ cm$^{-3}$ at the acceleration site for the left column pair, 
			and 3$\times$10$^{10}$ cm$^{-3}$ at the acceleration site for the right column pair.
			The ``down'' beam propagates to the left, toward the denser chromosphere, and the ``up'' beam to the right, towards the interplanetary medium.
			{\it Strong} refers to a flux of 2.7$\times$10$^{36}$ electrons/s, and {\it weak} to a flux of 1.0$\times$10$^{34}$ electrons/s.
			A 30 s accumulation time was used for the RHESSI simulated images.
		}
		\label{fig:rhessiSimImgs}
		\end{figure*}

	RHESSI simulated imaging (left column pair of Fig.~\ref{fig:rhessiSimImgs}) shows an elongated structure in the case of a {\it strong} ``up'' beam, but very little if the beam is {\it weak}.
	The situation improves only marginally in a denser corona (right column pair of Fig.~\ref{fig:rhessiSimImgs}): 
	the emissivity is indeed increased, but the spatial extend of the emission region is decreased, to a more localized source, as the column density traversed by the beam is thicker. 
	Going to higher energies is usually of no help, as the fluxes are much smaller (see also Appendix~\ref{appendix:dIdN}).
	It is easier to associate an elongated source as coming from a beam of electrons:
	a more compact source is more easily associated to an acceleration region or plasmoid with trapped particles.
	FOXSI, with its far better dynamic range (50'' away from the main source, its sidelobes are at the 10$^{-3}$ level, as opposed to $\approx$0.1 with RHESSI, S. Christe, PhD Thesis), is much better equipped than RHESSI (typical dynamic range of $\approx$10) to discriminate between these two cases.
	
	The ``up'' beam can clearly be imaged (as an elongated structure) by RHESSI if it is {\it strong}, and if the ``down'' component is {\it weak} or non-existent (occulted).
	No such observations have been reported so far, leading to the conclusion that ``up'' beams may very well always be of the {\it weak} kind, i.e. with fluxes $\lesssim$10$^{34}$ (electrons above 10 keV)/s.
	It also shows that in partially disk-occulted events, coronal emission produced by a {\it strong} beam in a flare loop is easily observable by RHESSI \citep{Krucker2007a,Krucker2008}.

	Because of instrumental sidelobes, the presence of a non-occulted thermal coronal source, such as typically produced by a flare loop, 
	will completely mask any beams, even if spatially separated, in the case of RHESSI.
	Even FOXSI's much-reduced sidelobes will only marginally allow it to observe {\it strong} beams, while {\it weak} beams most assuredly not
	(for quick comparison with top plots of Figure~\ref{fig:sp_prof_Eco10}: the thermal flux generated at 10 keV by a typical flare-like 10 MK, 10$^{49}$ cm$^{-3}$ source is about 3$\times$10$^5$ photons s$^{-1}$ cm$^{-2}$ keV$^{-1}$).

\subsection{GOES response}

	Table \ref{tab:goesresp} displays the expected response of GOES (GOES 10, both X-ray channels) to the {\it non-thermal} bresstrahlung from our beams of electrons.
	Were the ``up'' beams of the {\it strong} kind, GOES should easily observe them.
	In case of {\it weak} ``up'' beams, the emission can easily go unnoticed: it lies beneath the digitization level of the instrument (about A0.3 level).

	For information, Table \ref{tab:goesTEM} lists the temperature and emission measures that would be derived from the fluxes of Table \ref{tab:goesresp},
	using {\it Solarsoft}'s two-filter ratio method. 
	This method assumes an isothermal plasma, and indicates temperatures of 20--23 MK.
	The emission measure results scale well with beam fluxes ({\it weak} being 0.0037 times the {\it strong}) for the Mewe code.
	For the Chianti code, there is a software limitation due to the fact that the code is not meant to deal with such low photon fluxes, resulting in inconsistent numbers:
	they were hence not displayed for the case of the {\it weak} beam.
	For comparison, \citet{Feldmann1996} find in their statistical study that solar flares with such high ($\approx$22 MK) temperatures have an X-ray class of about M3.0, 
	i.e. emission from both {\it strong} (A4.4) and {\it weak} (A0.02), ``up'' coronal beams, if observed, would stand apart on a (temperature) vs. (GOES X-ray class) plot 
	(neglecting any local heating by the beams; see Section~\ref{sect:beamheating} for a discussion).

\subsection{Hinode/XRT response}

\begin{figure*}[ht!]
		\centering
		\includegraphics[width=11.66cm]{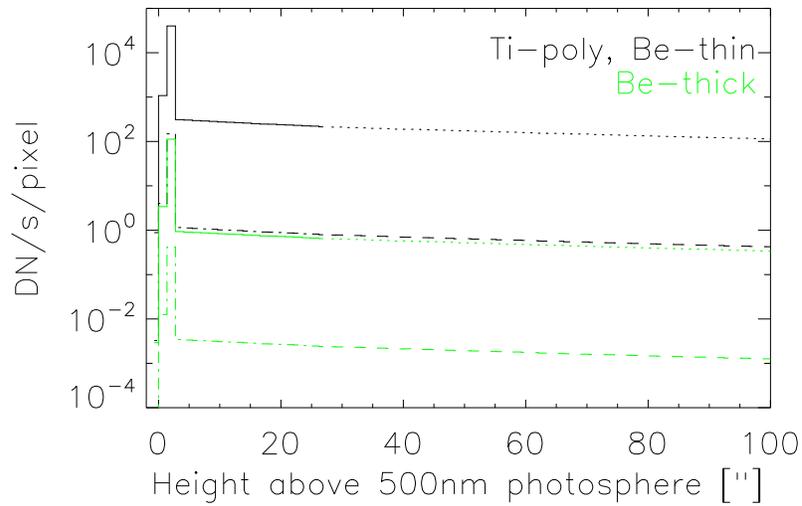}
		\caption{
			Spatial profile of ``up'' and ``down'' electron beams, convoluted with two different Hinode/XRT filters: Be-thin (very similar to Ti-poly) in {\it black}, and Be-thick in {\it green}.
			{\it Solid line:} Strong ``down beam'', {\it dotted line:} strong ``up'' beam, {\it dashed line:} weak ``up beam'',  {\it dot-dashed line:} weak ``down beam''.
			The acceleration site is located 30'' above the photosphere.
			The fact that the Be-thin response to a {\it weak} beam is similar to the Be-thick response to a {\it strong} beam is a coincidence.
		}
		\label{fig:XRT_prof}
		\end{figure*}

\begin{figure*}[ht!]
		\centering
		\includegraphics[width=15cm]{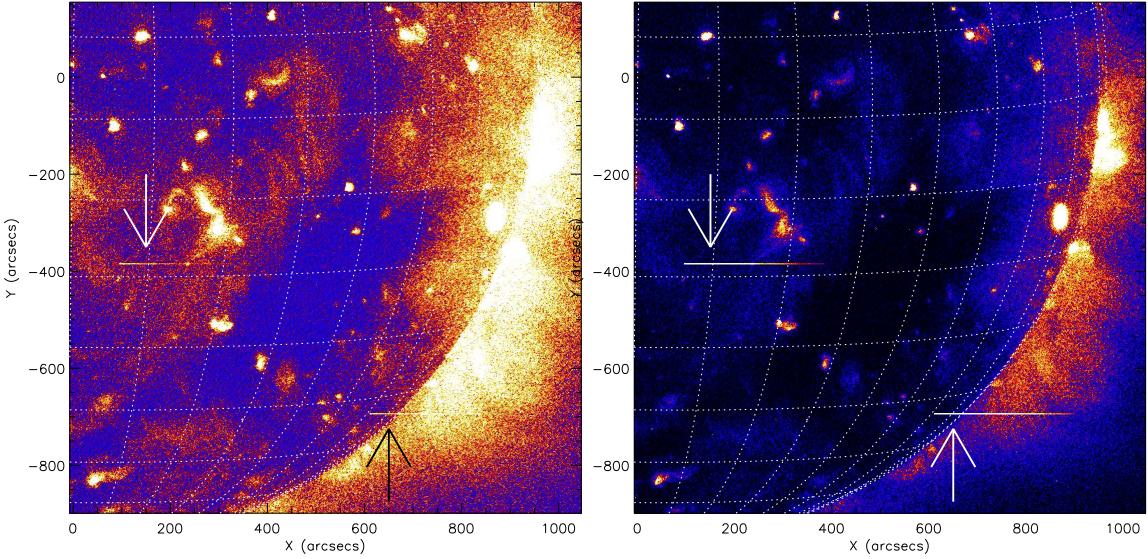}
		\caption{
			XRT Be-thin images (calibrated with the Solarsoft routine xrt\_prep.pro) taken on 2007/10/25 00:00:24 (32.76 s exposure), a relatively quiet time (GOES X-ray level constant at the A7.3 level),
			with non-thermal X-ray emission (convoluted with the filter response) from two hypothetical electron beams added.
			The electron beams start at around (600,-700) and (100,-380), propagate to the right perpendicularly to the line of sight,
			have a diameter of 3'', and have been truncated 300 Mm beyond their origin.
			{\it Left:} Electron beams are of the {\it weak} kind (i.e. 10$^{34}$ electrons/s above 10 keV, over the 32.76s exposure time),
			{\it Right:} Electron beams are 10 times the {\it weak} kind (i.e. 10$^{35}$ electrons/s, over the 32.76s exposure time),
			The basic image pixels span values between about 0 and 20 DN/s, but for easier identification of the beam features, the color scale dynamic range have been chosen as follows:
			{\it left:} 0 to 0.4;
			{\it right:} 0 to 1.
		}
		\label{fig:BethinSimEbeam}
		\end{figure*}

	The new Hinode/XRT \citep{Golub2007} instrument has a whole suite of different filters. 
	Figure~\ref{fig:XRT_prof} shows the spatial profile from coronal electron beams observed with some of them.
	As can be seen in Figure~\ref{fig:BethinSimEbeam}, a {\it weak} ``up'' electron beam is observable with XRT, 
	provided that a careful choice of image color scale is made.

	Optimal conditions are:
	\begin{itemize}
		\item Long exposures ($>$30 secs)
		\item Thin filter (such as Be-thin), for their better spectral response
		\item Usage of appropriate image color scales, allowing for certain weak features to be revealed:
			even a weak beam-like feature can be distinguishable from the image noise or other dominant features because of its structure in the image (e.g. a straight line,...).
	\end{itemize}
\subsection{Effect of beam heating} \label{sect:beamheating}

	This short section tries to estimate the effect of heating of the local corona by the non-thermal ``up'' beam:
	can the resulting X-ray thermal emission be greater than the X-ray non-thermal emission?
	
	The non-thermal power lost in collisions can in principle heat the local medium: 
	the amount of non-thermal power dumped along each path element can be calculated.
	Given a beam duration $\Delta t$, beam area, and in the absence of thermal losses (from either heat conductivity or thermal radiation), an upper limit to the temperature increase for the ambiant corona along the path of the beam can be determined
	(mathematical details in Appendix \ref{appendix:openloopheating}).
	
	As a consistency check on the assumption of no thermal losses, the timescale for heat conductivity losses can be very roughly estimated (see details in Appendix \ref{appendix:closedloopheating}),
	using the density scale height of the heated medium. 
	If this timescale is shorter than the beam duration, then heat losses should be included.
	The treatment presented in Appendix \ref{appendix:closedloopheating} is best applied to a closed loop system, but was used as proxy for our case with an open field line,
	taking the loop length $L$ to be the density scale height. 
	In this treatment, the resulting temperature depends weakly on $L$, and our model was deemed sufficiently accurate to provide a rough upper limit of ambiant temperature.

\begin{figure}[ht!]
		\centering
		\includegraphics[width=7.5cm]{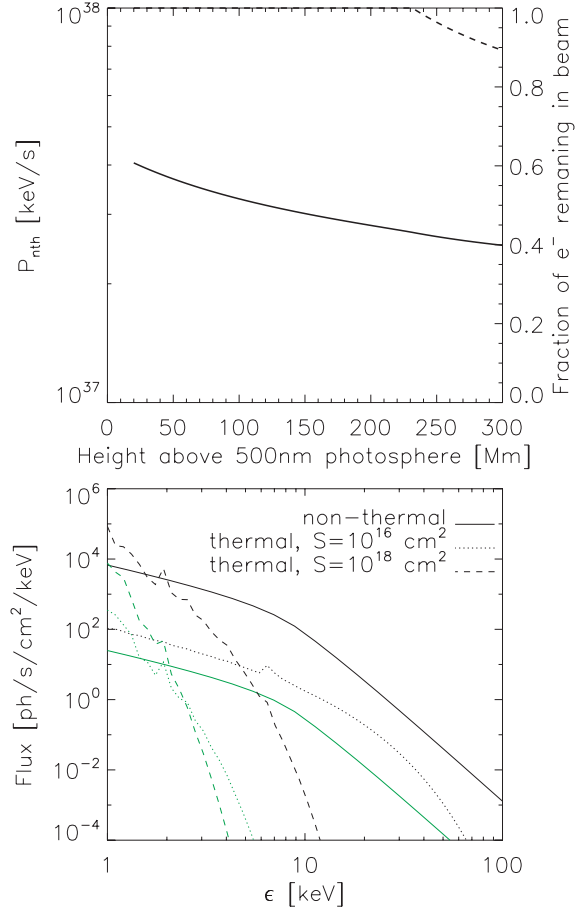}
		\caption{
			{\it Top}: Profile of the non-thermal power left in a {\it strong} ``up'' beam, as it propagates through the corona.
				Scale by 0.0037 for the {\it weak} beam counterpart.
			{\it Bottom}: Resulting non-thermal and estimated thermal emissions (see Section~\ref{sect:beamheating}).
			Two different beam sections $S$ were used to evaluate the plasma thermal response. 
			{\it Solid} line are the non-thermal spectra.
			The {\it dotted} (about 1" beam radius) and {\it dashed} (about 10" beam radius) lines are thermal spectra determined from beam coronal heating.
			{\it Black} pertains to the {\it strong} beam, {\it green} to the {\it weak} beam.
		}
		\label{fig:beam_heating}
		\end{figure}

	Fig.(\ref{fig:beam_heating}) ({\it bottom}) displays the thermal spectrum expected from our non-thermal beams of electrons, 
	for both {\it weak} and {\it strong} cases, as well as for different beam areas: (10$^{16}$ cm$^2$ is about 1" radius, and (10$^{18}$ cm$^2$ is for a beam with a radius of about 10" radius.)
	In the case of a {\it weak} beam, both conduction and radiative loss timescales were found to be greater than the duration of the beam's injection ($\Delta t$=100 s).
	For the {\it strong} beam with a small section ({\it black dotted} line in Fig. \ref{fig:beam_heating}), the heat conduction energy loss timescale (Eq. \ref{eq:tcond}, taking $L$ to be the barometric scale height $H_n$=10$^{10}$ cm) was found to be many orders of magnitude smaller than $\Delta t$,
	leading to the use of Eq. (\ref{eq:Teq}) to better estimate the plasma temperature. Table \ref{tab:beamheatingTEM} summarizes the temperatures and emission measures derived.

	In the case of a {\it weak} ``up'' beam, it appears that the thermal emission should be negligible in comparison to the beam's non-thermal emission, 
	at RHESSI energies ($>$3 keV).
	At energies below $\sim$3 keV, it is the thermal emission that is expected to dominate.
	This result is not very sensitive to the position of the low-energy cutoff (all else, including the total electron flux, being equal).
	
	A {\it strong} ``up'' beam with wide cross-section behaves much the same as the {\it weak} beam cases.
	A {\it strong} ``up'' beam with small cross-section, while very hot, is masked by the even stronger non-thermal emission.

	To summarize, non-thermal emission are expected to prevail at energies above $\sim$3 keV.
	Below $\sim$3 keV, thermal emission is likely to dominate.

	Both XRT on {\it Hinode} and SXT on {\it Yohkoh} observe below $\sim$2 keV.
	This raises the very interesting possibility that the SXR emission from X-ray jets that are observed by XRT
	and previously by SXT, to which Type III radio bursts have been sometimes associated \citep{Shibata1992, Aurass1994, Kundu1995, Raulin1996}, 
	could very well be the direct result of such heating.
	X-ray jets on the Sun would then be expected to occur whenever coronal and interplanetary electron beams are observed.
	The temperatures (5 MK) and emission measures (10$^{44}$ cm$^{-3}$) obtained by \citet{Kundu1995} for their X-ray coronal jet are qualitatively near those of 
	Table \ref{tab:beamheatingTEM} for the case of the {\it weak} beam with small cross-section, futher supporting this claim.

	A careful search of observations for spatially and temporally correlated HXR, radio Type III, and X-ray jets has been initiated.
	
\section{Upper limits from observations}

	This section includes some observational facts to our discussion so far.
	The first part deals with the ``coronal beam associated with X-ray jets'' aspect that was suggested earlier,
	followed by a brief discussion on constraints imposed by the oft-observed lack of  correlation between 
	Type III radio bursts and X-ray emission.

\subsection{X-ray jets as coronal beams}
	
	As the previous discussion has suggested that X-ray jets might be associated with coronal electron beams,
	we have examined a few polar X-ray jets \citep{Cirtain2007} observed with {\it Hinode} XRT long-exposure images.
	Nine have been observed in the period 2007/03/11 21:00 to 2007/03/12 06:00 UT, near the solar limb (but not occulted).
	In none of these cases were any X-ray flux enhancements observed with RHESSI, or GOES:
	
	\begin{itemize}
		\item 
			Table~\ref{tab:lcdetection} gives the amount of electron that an {\it up} beam must contain in order to be detectable in spatially-integrated 
			RHESSI Observing Summary countrates \citep{Schwartz2002}.
			The absence of any clear observation puts the upper limit on the total number of electrons in a coronal beam to 0.6$\times$10$^{35}$ electrons above 10 keV
			(3--$\sigma$ detection level) over short timescales of a few seconds, and 4.3$\times$10$^{35}$ electrons on timescales of a few minutes.

		\item 
			The GOES low and high channels remained also flat. 
			For detection by visual inspection of the GOES lightcurves, an increase of $\approx$2$\times$10$^{-9}$ W/m$^2$ in the low (1--8$\AA$) channel
			was required over a 3 s time bin (rough estimate).
			This corresponds (see Table~\ref{tab:goesresp}) to 6$\times$10$^{35}$ electrons above 10 keV
			(1.5$\times$10$^{36}$ electrons in the high (0.5--4$\AA$) channel).

		\item One of the X-ray jets (2007/03/12 05:18 UT) lasted about five minutes.
			RHESSI imaging over this five minute interval yields no reliable image in the 3--6 or 6--12 keV bands.
			As any RHESSI source typically needs at least $\approx$300 counts/detector (empirical value) to be successfully characterized, 
			this translates into a needed count rate of approximatively 1 counts/s/detector over that time interval.
			This puts the needed number of electrons for good imaging to
			1.2$\times$10$^{37}$ electrons above 10 keV for the 3--6 keV band, and 3$\times$10$^{36}$ electrons for the 6--12 keV band. 
	\end{itemize}

	Overall, RHESSI lightcurves are more sensitive than GOES lightcurves, and further indicate that observed X-ray jets did not expel more than 0.6$\times$10$^{35}$ electrons above 10 keV
	on timescales of a few seconds, and no more than $\approx$5$\times$10$^{35}$ electrons above 10 keV over timescales of a few minutes.
	From inspection of Fig.~\ref{fig:sp_prof_Eco10} ({\it bottom left}), about 10$^{33}$ electrons is required for FOXSI imaging.

	In cases were the acceleration site is at lower densities than our start density of 3$\times$10$^9$ cm$^{-3}$ (a typical high value),
	then these upper limits get proportionally higher. For example, were the start density 3$\times$10$^{8}$cm$^{-3}$, 
	the upper limit on the number of accelerated electrons is increased tenfold.

\subsection{Type III radio bursts from coronal beams}
		
	The detection thresholds from the previous section can be used again:
	For X-ray detection through RHESSI lightcurves, a coronal electron beam would require about 0.6$\times$10$^{35}$ electrons above 10 keV over 4 s,
	or about 4$\times$10$^{35}$ electrons above 10 keV over a few minutes.
	RHESSI characterization by imaging requires at least 3$\times$10$^{36}$ electrons above 10 keV.

	The fact that no clear spatial and temporal correlation of Type III radio burst and non-thermal X-ray emission beyond the limb has ever been established
	is already an indicator that electron beams must have typically less than these numbers of electrons.
	A systematic search using data from the Nan\c{c}ay Radioheliograph and RHESSI will be initiated shortly.
	The best case so far of such an event has been discussed in \citet{Krucker2008}, and mentionned briefly in the conclusion.
	
\section{Summary and Conclusion}

	(1) {\it Strong} escaping (``up'') beams, i.e. with fluxes comparable to the usual chromospheric HXR-producing flare electrons ($\gtrsim$10$^{38}$ electrons above 10 keV),
	should easily be detectable and imageable with RHESSI, provided any chromospheric footpoint is occulted.
	The absence of such observations supports the scenario established so far, i.e. that escaping electrons are fewer in number than a few tenths of a percent of those hitting the chromosphere.
	This in turns hints at asymmetries in the overall standard acceleration scenario.
	Possible explanations can range from 
		(a) the presence of a ``collapsing trap'' mechanism \citep[as described e.g. in][]{Karlicky2004} that enhances the number of accelerated flare electrons, 
		but not the escaping electrons on open field lines,
		(b) the possibility that the main acceleration actually takes place elsewhere than in the high corona, such as in the footpoints, as suggested by \citet{Fletcher2008},
		(c) the possibility that escaping electron beams are a secondary energy release phenomenon, triggered by electromagnetic waves from the flare electrons \citep{Sprangle1983},
		or (d) the presence of a secondary reconnection process higher up in the corona, where particle densities are much lower, connecting to open field lines \citep[see e.g.][]{Vrsnak2003}.
	
	(2) GOES is not expected to observe anything of note from {\it weak} beams (beams with $\lesssim$10$^{36}$ electrons above 10 keV).
	Escaping {\it weak} beams appear to be just below RHESSI's imaging capabilities (even with footpoints occulted), 
	marginally within Hinode/XRT's imaging capabilities, but well within FOXSI's.
	For XRT, thin filters and long exposures are required, as is a careful choice of image dynamic range.
	A systematic search of the XRT data, particularly those with long exposure times is currently underway:
	For the year 2007, XRT was observing near the solar limb (partial disk images, with image center $>$600" from Sun center) with thin filters and long exposures ($>$30 s) 0.65\% of the time.
	NOAA\footnote{ftp://ftp.ngdc.noaa.gov/STP/SOLAR\_DATA/SOLAR\_RADIO/SPECTRAL/SPEC\_NEW.07} reports about 500 different Type III bursts during 2007 (a very quiet year).
	Assuming they were produced by beams of electrons that were 30 seconds long, it means that Type III-producing electron beams occur 0.05\% of the time.
	The probability for simultaneous occurence of a Type III burst and XRT long-exposure observation with a thin filter is hence p$\approx$3$\times$10$^{-6}$.
	Between the start of the Hinode mission, and the end of January 2008, about n=6000 long-exposure pictures with thin filters were taken by XRT.
	The chance of there being at least one electron beam caught within that sample can be hence estimated to be 1-(1-p)$^n$ $\approx$ np $\approx$ 2\%,
	i.e. we probably haven't observed it.
	
	(3) Coronal emission due to beam heating is not strong enough to mask the non-thermal bremsstrahlung emission of the upgoing beam at energies above 3 keV (i.e. in the enrgy ranges of RHESSI and FOXSI).
	With an instrument such as XRT, a rough estimation leads us to expect that the thermal emission might indeed mask the non-thermal emission.

	(4) We have raised the possibility that SXR jets might be the result of local heating by the propagating coronal/interplanetary electron beams.
	This is consistent with the fact that interplanetary electron beams have been found to be temporally correlated with SXR plasma jets \citep{Wang2006,Pick2006,Nitta2008}.
	On the other hand, no non-thermal HXR emission has ever been spatially associated with those SXR jets, probably due to lack of sensitivity.

 	(5) The absolute minimum amounts of electrons needed for X-ray {\it detection} and for {\it characterization through imaging} are (assuming optimal start densities and minimal backgrounds):
	\begin{itemize}
		\item	$\gtrsim$10$^{35}$ electrons above 10 keV: for detection (and localisation via coarse imaging, to the $\approx arc minute level$) by RHESSI
		\item	$\gtrsim$6$\times$10$^{35}$ electrons above 10 keV: for detection by GOES
		\item   $\gtrsim$3$\times$10$^{35}$ electrons above 10 keV: for imaging by Hinode/XRT
		\item	$\gtrsim$3$\times$10$^{36}$ electrons above 10 keV: for imaging by RHESSI (with sufficient statistics to observe structures to the $\approx$10'' level)
		\item	$\gtrsim$10$^{33}$ electrons above 10 keV: for imaging by FOXSI (180 cm$^2$ effective area detector, assumes zero background)
	\end{itemize}

	Appendix~\ref{appendix:requirements} is a list of optimal observations for identification and characterization of escaping coronal electron beams.
				
	An order of magnitude estimate on the expected number of beams with enough electrons to be characterized through RHESSI X-ray imaging can be done as follows:
	Assuming that $\approx$10\% of the $\approx$120 electron events per year that WIND observes around solar maximum produce $\approx$10$^{34}$ electrons above 22 keV, 
	or $\approx$10$^{35}$ electrons above 10 keV, this leads to $\approx$12 events with $\gtrsim$10$^{35}$ electrons above 10 keV per year.
	Using the 1.4 power-law negative spectral index found in peak interplanetary electron flux distributions (P.H. Oakley, {\it priv. comm.}), 
	this leads to an estimate of about 0.5 interplanetary beams with $\gtrsim$10$^{36}$ electrons above 10 keV per solar-maximum year.
	With three spacecrafts (WIND, STEREO A \& B), the expectation becomes 1.5 per solar-maximum year.
	Periods when occulted flares can be observed from Earth and a spacecraft with in situ intruments magnetically connected to its escaping particles
	are around the middle of 2009 with STEREO A and around 2014 with STEREO B.
	
	The best case published so far of an observation of a coronal electron beam, using RHESSI and XRT data, has been discussed in Krucker et al. (2008).
	Yet, the XRT coverage was not optimal, no radio imaging was available, and, most importantly, 
	the number of electrons in the interplanetary beam was at least an order of magnitude below what was inferred from the coronal X-ray emission 
	(3$\times$10$^{33}$ vs. 10$^{34}$--3$\times$10$^{36}$ electrons above 20 keV).
	The author argue that the in-situ measurements, which sample only a very small fraction of the beam's breadth, 
	might make erroneous assumptions on its spatial distribution, and that in reality many more escaping electrons could be present.
	
	With the latest additions to the fleet of sun-observing spacecraft (STEREO, HINODE, SDO) and the solar activity rising, 
	it is expected that several such events will be sufficiently observed. 
	Future spacecraft missions in the inner heliosphere (Solar Orbiter, Sentinels, Solar Probe) will provide regularly such observations at much higher sensitivity.
	The scheduled 5-minute FOXSI rocket mission will have the dynamic range and sensitivity required to images faint X-ray emission from outgoing electron beams, 
	but the chance of observing a radio type III burst during a 5 minute flight is close to zero. 
	A future space mission with a focusing optics telescope dedicated to solar observations, however, would revolutionize our understanding of electron acceleration in solar flares. 
	The Nuclear Spectroscopic Telescope Array (NuSTAR) Small Explorer satellite \citep{Harrison2005}, to be launched in 2011, uses HXR focusing optics for astrophysical observations with an effective area of 1000 cm$^2$. 
	With a solar mission of similar size as NuSTAR, HXR emission from escaping electron beams will be generally detected with excellent statistics allowing us to spectrally image the electron acceleration region and trace electron beams from their acceleration region down to the chromosphere as well as into interplanetary space.

\acknowledgments

This work was supported by NASA Heliospheric GI awards NNX07AH74G and NNX07AH76G, and by Swiss National Foundation (SNSF) grant PBEZ2-108928.
We would like to thank the anonymous referee for his constructive comments.
{\it Facilities:} \facility{RHESSI}, \facility{Hinode (XRT)}, \facility{GOES}, \facility{FOXSI}.



\appendix
\section{X-ray spatial emission profiles}	\label{appendix:dIdN}

	Using Kramers' simplified differential bremsstrahlung cross-section $Q(\varepsilon, E)=\bar{Z}^2\frac{\kappa_{BH}}{\epsilon E}$ in Eq. (\ref{eq:int}), one arrives at:
	
		\begin{equation}
			\frac{dI}{dN}(\varepsilon,N) = (\delta-1) \frac{F_1}{E_1} \bar{Z^2} \frac{\kappa_{BH}}{8\pi D^2} \frac{1}{\varepsilon} \left( \frac{2KN}{E_1^2} \right)^{-\delta/2}  \, B \left( \frac{1}{1+u},\frac{\delta}{2},\frac{1}{2} \right)
		\end{equation}
	Where  $\bar{Z^2}\approx$1.44 in the corona, $\kappa_{BH}$=$\frac{8}{3} \, \alpha \, r_e^2 \, m_ec^2$=7.9$\times$10$^{-25}$ cm$^2$ keV, $B(x,a,b)$ is the incomplete beta function, and:
		\begin{equation}
			u = \frac{1}{2KN} \,\,\, \max( \varepsilon^2, E_1^2 - 2KN )
		\end{equation}
	Similar to Brown, 2002, except that we provide for the emission below the low-energy cutoff.

\begin{figure}[ht!]
		\centering
		\includegraphics[width=11.66cm]{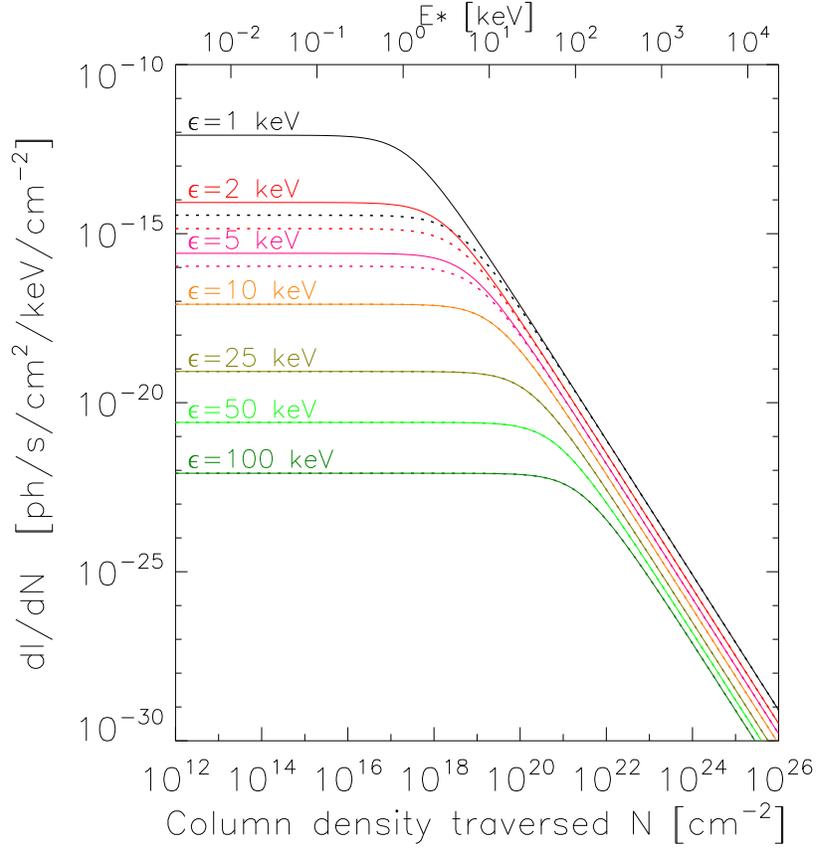}
		\caption{
			Photon flux per unit column density, as a function of traversed column density $N$.
			The initial electron beam had $\delta$=4, and 2.7$\times$10$^{36}$ (electrons above 10 keV, per second).
			{\it Solid lines:} no cutoff, {\it dotted lines:} low-energy cutoff at 10 keV.
			$E_*=\sqrt{2KN}$, with $K=2.6\times 10^{-18}$ cm$^2$ keV$^2$, is the initial electron energy that a column density $N$ brings to a stop.
		}
		\label{fig:dI_dN}
		\end{figure}

	In the absence of low-energy cutoff ($E_1$=0), or at least as long as $\varepsilon^2 \ge E_1^2-2KN$, we have:
		\begin{eqnarray}
			\frac{dI}{dN}(\varepsilon,N)	& \approx (\delta-1)\frac{F_1}{E_1} \bar{Z^2} \frac{\kappa_{BH}}{8\pi D^2} E_1^{\delta} &	\left\{	\begin{array}{ll} 
										\varepsilon^{-1} \cdot (2KN)^{-\delta/2} \cdot B(\frac{\delta}{2},\frac{1}{2}) &, \,\,\, u \ll 1	\\
										\varepsilon^{-\delta-1}	&, \,\,\, u \gg 1	\\
										\end{array} \right.
		\end{eqnarray}
	where $B(a,b)$ is the Beta function.
	I.e. the emissivity at a certain photon energy $\varepsilon$ is constant along the path of the beam, until $u$ decreases to $\approx$1, i.e. $\varepsilon \approx \sqrt{2KN}$, after which it falls rapidly with increasing $N$, as shown in Fig.~\ref{fig:dI_dN}.

\begin{figure}[ht!]
		\centering
		\includegraphics[width=11.66cm]{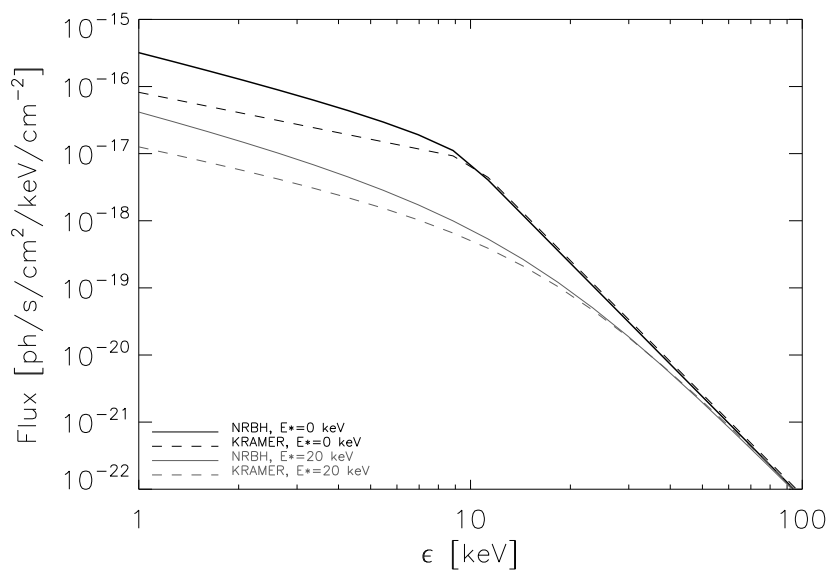}
		\caption{Photon flux per unit column density, for the typical strong beam ($\delta$=4, $E_1$=10 keV, $F_1$=2.7$\times$10$^{36}$ e$^-$/s), at injection (N=0, {\it black}),
			and after it has traversed a column density N=8$\times$10$^{19}$ cm$^{-2}$ ({\it gray}, stopping all electrons of initial energies below $E_*$=20 keV.). 
			{\it Solid} lines: using the non-relativistic Bethe-Heitler (NRBH) cross-section,
			{\it Dashed} lines: using the Kramers cross-section.
			}
		\label{fig:comparison_NRBH_Kramer}
		\end{figure}

	Using the Kramers cross-section however does not yield a wholly accurate photon spectrum for $\varepsilon<E_1$ (Fig.~\ref{fig:comparison_NRBH_Kramer}).
	Hence, numerical evaluations of $\frac{dI}{dN}$ using Eq. (\ref{eq:int}) and the more proper non-relativistic Bethe-Heitler differential bremsstrahlung cross-section (3BN(a) of Koch \& Motz, 1959; Brown, 1971):
		\begin{equation}
			Q(\varepsilon,E) = \bar{Z^2} \frac{\kappa_{BH}}{\varepsilon E} \ln \left( \frac{1+\sqrt{1-\varepsilon/E}}{1-\sqrt{1-\varepsilon/E}} \right)
		\end{equation}
	needs to be used, in order to cover all possible cases.

\section{Number of non-thermal particles remaining in beam}	\label{appendix:Nnth}
		
		The number of electrons above reference energy $E_{ref}$ in a beam that has already traversed a column depth $N$ is given by:

		\begin{equation}
			N_{nth}(N) = \int_{E_{ref}}^\infty F(E,N) \, dE
		\end{equation}
	where $F(E,N)$ is as given by Eq. (\ref{eq:FEN2}), resulting in:
	
	\begin{equation} \label{eq:beam_remaining_electrons}
		N_{nth}(N)= F_1 E_1^{\delta-1} (\xi^2 + 2KN)^{(1-\delta)/2},
	\end{equation}
	where $\xi=\max ( \zeta, E_{ref})$. 
	
\begin{figure}[ht!]
		\centering
		\includegraphics[height=8cm]{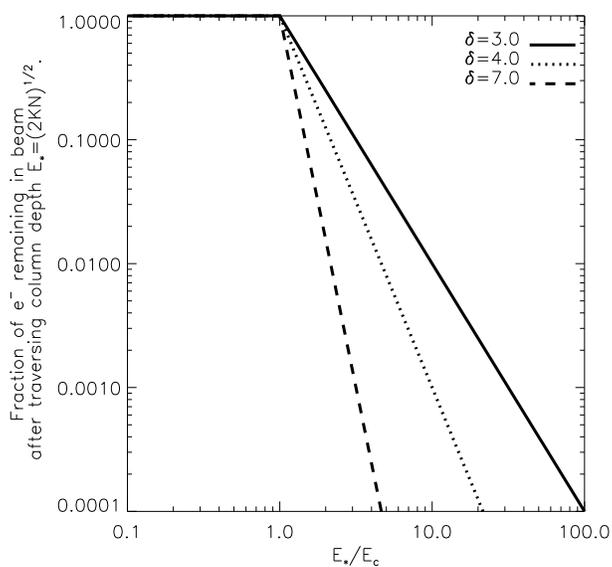}	
		\caption{
			Fraction of the intial number of electrons remaining in the beam, after a column density $N$ has been traversed, for different injected beam spectral indices $\delta$.
			$E_c$ is the injected beam's low-energy cutoff, and $E_*=\sqrt{2KN}$, with $N$ the column density traversed, and $K=2.6\times 10^{-18}$ cm$^2$ keV$^2$.
		}
		\label{fig:beam_fraction}
		\end{figure}

	For the case where $E_{ref}$=0 (i.e. all electrons in the beam), Eq. \ref{eq:beam_remaining_electrons} amounts to:
	
	\begin{eqnarray}
		N_{nth}(N)	& = &	\left\{	\begin{array}{ll} 
						F_1								& , \,\,\, \sqrt{2KN} \leq E_1	\\
						F_1 \cdot \left( \frac{E_1}{\sqrt{2KN}} \right)^{\delta-1}	& , \,\,\, \sqrt{2KN} \geq E_1	\\
						\end{array} \right.
	\end{eqnarray}
	As can be seen in Fig. (\ref{fig:beam_fraction}): a beam has lost {\it at least} 90\% of its electrons by the time it has traversed 3 times the column density required to stop electrons starting at its low-energy cutoff.

\section{Plasma heating by non-thermal beams of electrons: lossless case}	\label{appendix:openloopheating}
	The non-thermal power in a beam, at any point along its path, $P_{nth}(N)$ can be computed numerically:
		\begin{equation}
			P_{nth}(N)= \int_0^{\infty} E \cdot F(E,N) dE
		\end{equation}
	The non-thermal power loss per unit length $dz$, along the path of the beam, is computed as $-\frac{dP_{nth}}{dz}$,
	and is the same as the heat dumped per unit length in the corona along the path of the beam.
	Assuming no losses, the temperature of the corona, along the path of the beam, can be estimated to be:
	
		\begin{equation}	\label{eq:kT}	
			k_B T(z) = k_BT_0 + \frac{\Delta t}{3 \cdot S \cdot n_e(z)} \left( -\frac{dP_{nth}}{dz} (z) \right)	,
		\end{equation}
		where $S$ is the area of the beam, $\Delta t$ is the duration of the injection, $n_e(z)$ the local electron density, and $T_0\approx$2MK the initial temperature of the corona.

	The differential emission measure, along the path of the beam, is:
	
		\begin{equation}	\label{eq:dEM}
			\frac{dEM(z)}{dz} = n_e^2(z) \cdot S
		\end{equation}

	Both Eqs.~(\ref{eq:kT}) and (\ref{eq:dEM}) can be used to compute the total thermal spectrum generated by beam heating, by integrating over the whole path of the beam (or at least the portion within the instrument's field of view).
	This has been done in Section \ref{sect:beamheating}.
\section{Plasma heating by non-thermal beams of electrons: scenario with conductive losses}	\label{appendix:closedloopheating}

	The problem is carried a little bit further than as in Appendix \ref{appendix:openloopheating}:
	the energy flux due to heat conduction in the direction parallel to the magnetic field is:
	\begin{equation}
		j_Q = \kappa \frac{dT}{dz} ,
	\end{equation}
	where $\kappa= \alpha T^{5/2}$ keV s$^{-1}$ K$^{-1}$ cm$^{-1}$ is the Spitzer conductivity \citep{Spitzer1962,BenzBook}, with $\alpha$=537.5 keV s$^{-1}$ cm$^{-1}$ K$^{-7/2}$.
	Heat conduction perpendicular to the magnetic field is negligible.
	
	Replacing $j_Q$ by $\frac{1}{S}\frac{dE_{th}}{dt} \approx \frac{1}{S} \frac{E_{th}}{\tau_{cond}} \approx \frac{1}{S} \frac{3k_BTn_eV}{\tau_{cond}} $, and $\frac{dT}{dz}$ by $\frac{T}{L}$, where $L$ ans $S$ are respectively the length and section of the flux tube, $V=SL$ is the volume, $n_e$ the average electron density, $k_B$ Boltzmann's constant, one finds the heat conductivity energy loss time-scale:
	\begin{equation}	\label{eq:tcond}
		\tau_{cond} = \frac{3 k_B}{\alpha} \frac{n}{T^{5/2}} L^2
	\end{equation}

	In a volume $V=SL$, at equilibrium, the amount of thermal energy loss due to heat conductivity is the same as the amount of non-thermal power dumped:
	
	\begin{equation}
		P_{nth}	= \frac{E_{th}}{\tau_{cond}}
	\end{equation}
	(assuming $\tau_{cond} \ll \tau_{rad}$, the radiative loss timescale, which is usually the case in hot flare plasmas near the impulsive phase of the flare).
	Hence:	
		\begin{equation} \label{eq:Teq}
			T_{eq} = \left( \frac{P_{nth}}{\alpha} \frac{L}{S} \right)^{2/7}
		\end{equation}

	Notice that this equilibrium temperature $T_{eq}$ is independent of density and filling factor, and that a factor 2 error on any one parameter translates into a 22\% error in $T_{eq}$.
	For example, taking $P_{nth}$=$\frac{\delta-1}{\delta-2} F_1 E_1$= 6.5$\times$10$^{28}$ erg/s=4$\times$10$^{37}$ keV/s, $S$=10$^{17}$ cm$^2$, and $L$=10$^9$ cm, one finds $T_{eq}$=47.6MK, 
	or $k_BT_{eq}$=4.2 keV.
\section{Requirements for unambiguous identification of X-ray emission from electron beams flowing into interplanetary space}	\label{appendix:requirements}

	The following is a list of requirements that should be ideally fulfilled in order to be able to find and characterize such beams:
	\begin{itemize}
		\item {\it Minimum number of electrons over 10 keV (typically over a $\sim$10--1000 s accumulation interval):} 
		RHESSI: 3$\times$10$^{36}$; 
		Hinode/XRT: 3$\times$10$^{35}$ (assuming the spectral contribution of beam heating is negligible); 
		FOXSI: 10$^{33}$ (assuming zero background).
		These numbers are for very careful examintation of the data. 
		Multiply these requirements by $\approx$3 for casual searches.
		
		\item {\it X-ray emission above the solar limb}: 
			Projection effects are expected to be less important, allowing an easier characterization of observed features.
			Moreover, the flare footpoint emission from the ``down'' beam, if any, should be occulted, 
			to limit its masking effects on the weaker emission from the electron beam.
			This last requirement is not as important for FOXSI as it is for RHESSI: FOXSI's sidelobes are below the 10$^{-3}$ level 50'' away from the main source (S. Christe, PhD thesis), as opposed to $\approx$0.1 for RHESSI.
		\item {\it Elongated X-ray image}: 
			If observing at the limb, and assuming a more-or-less radial propagation, X-ray imaging should display an elongated source in the radial direction.
			More compact sources could be assimilated to other origins, such as a plasmoid, or even a current sheet \citep{Bemporad2006}.
			Ideal acceleration site density for RHESSI is around 3$\times$10$^9$ cm$^{-3}$: this ensures source elongation at the energies where RHESSI is the most sensitive (6--10 keV).
		\item {\it Non-thermal X-ray spectrum}: 
			The X-ray emission should have a non-thermal spectrum, from which beam characteristics (spectral index and particle number) can be extracted and compared to in situ measurements.
		\item {\it In situ electron spectrum}: 
			An in situ instrument such as on board WIND or STEREO should be magnetically connected to the flare, in order for them to detect the expelled electron beams.
			The extracted electron beam characteristics (spectral index, particle number) shoud be compared to those deduced from X-ray emission \citep[as in e.g.][]{Lin1971,Krucker2007}.
		\item {\it Metric/Decimetric Type III radio emission, co-temporal and co-spatial with the non-thermal X-ray emission}:
			While not absolutely necessary, tracking an escaping electron beams's Type III radio emission 
			\citep[as was done with the Nancay Radioheliograph in e.g.][]{Paesold2001}
			would of course strengthen the case.
			The upcoming FASR \citep[Frequency Agile Solar Radiotelescope][]{Bastian2003} will provide such information with unprecedented coverage.
			At lower frequencies, LOFAR\footnote{http://www.lofar.org} could be used.
	\end{itemize}

\bibliographystyle{apj}
\bibliography{psh_biblio}

\begin{thebibliography}{47}
\expandafter\ifx\csname natexlab\endcsname\relax\def\natexlab#1{#1}\fi

\bibitem[{{Aurass} {et~al.}(1994){Aurass}, {Klein}, \& {Martens}}]{Aurass1994}
{Aurass}, H., {Klein}, K.-L., \& {Martens}, P.~C.~H. 1994, \solphys, 155, 203

\bibitem[{{Bastian}(2003)}]{Bastian2003}
{Bastian}, T.~S. 2003, Advances in Space Research, 32, 2705

\bibitem[{{Bastian} {et~al.}(1998){Bastian}, {Benz}, \& {Gary}}]{Bastian1998}
{Bastian}, T.~S., {Benz}, A.~O., \& {Gary}, D.~E. 1998, \araa, 36, 131

\bibitem[{{Baumbach}(1937)}]{Baumbach1937}
{Baumbach}, S. 1937, Astron. Nach., 263, 121

\bibitem[{{Bemporad} {et~al.}(2006){Bemporad}, {Poletto}, {Suess}, {Ko},
  {Schwadron}, {Elliott}, \& {Raymond}}]{Bemporad2006}
{Bemporad}, A., {Poletto}, G., {Suess}, S.~T., {Ko}, Y.-K., {Schwadron}, N.~A.,
  {Elliott}, H.~A., \& {Raymond}, J.~C. 2006, \apj, 638, 1110

\bibitem[{{Benz}(1993)}]{BenzBook}
{Benz}, A.~O., ed. 1993, Astrophysics and Space Science Library, Vol. 184,
  {Plasma astrophysics: Kinetic processes in solar and stellar coronae}

\bibitem[{{Benz} {et~al.}(1982){Benz}, {Treumann}, {Vilmer}, {Mangeney},
  {Pick}, \& {Raoult}}]{Benz1982}
{Benz}, A.~O., {Treumann}, R., {Vilmer}, N., {Mangeney}, A., {Pick}, M., \&
  {Raoult}, A. 1982, \aap, 108, 161

\bibitem[{{Brown}(1971)}]{Brown1971}
{Brown}, J.~C. 1971, \solphys, 18, 489

\bibitem[{{Brown}(1972)}]{Brown1972}
---. 1972, \solphys, 26, 441

\bibitem[{{Brown} {et~al.}(2002){Brown}, {Aschwanden}, \& {Kontar}}]{Brown2002}
{Brown}, J.~C., {Aschwanden}, M.~J., \& {Kontar}, E.~P. 2002, \solphys, 210,
  373

\bibitem[{{Christe} {et~al.}(2008){Christe}, {Krucker}, \& {Lin}}]{Christe2008}
{Christe}, S., {Krucker}, S., \& {Lin}, R.~P. 2008, \apjl, 680, L149

\bibitem[{{Cirtain} {et~al.}(2007){Cirtain}, {Golub}, {Lundquist}, {van
  Ballegooijen}, {Savcheva}, {Shimojo}, {DeLuca}, {Tsuneta}, {Sakao}, {Reeves},
  {Weber}, {Kano}, {Narukage}, \& {Shibasaki}}]{Cirtain2007}
{Cirtain}, J.~W., {Golub}, L., {Lundquist}, L., {van Ballegooijen}, A.,
  {Savcheva}, A., {Shimojo}, M., {DeLuca}, E., {Tsuneta}, S., {Sakao}, T.,
  {Reeves}, K., {Weber}, M., {Kano}, R., {Narukage}, N., \& {Shibasaki}, K.
  2007, Science, 318, 1580

\bibitem[{{Dennis}(1985)}]{Dennis1985}
{Dennis}, B.~R. 1985, \solphys, 100, 465

\bibitem[{{Dennis} {et~al.}(1984){Dennis}, {Benz}, {Ranieri}, \&
  {Simnett}}]{Dennis1984}
{Dennis}, B.~R., {Benz}, A.~O., {Ranieri}, M., \& {Simnett}, G.~M. 1984,
  \solphys, 90, 383

\bibitem[{{Dulk}(1985)}]{Dulk1985}
{Dulk}, G.~A. 1985, \araa, 23, 169

\bibitem[{{Elwert} \& {Haug}(1971)}]{Elwert1971}
{Elwert}, G., \& {Haug}, E. 1971, \solphys, 20, 413

\bibitem[{{Fainberg} \& {Stone}(1974)}]{Fainberg1974}
{Fainberg}, J., \& {Stone}, R.~G. 1974, Space Science Reviews, 16, 145

\bibitem[{{Feldman} {et~al.}(1996){Feldman}, {Doschek}, {Behring}, \&
  {Phillips}}]{Feldmann1996}
{Feldman}, U., {Doschek}, G.~A., {Behring}, W.~E., \& {Phillips}, K.~J.~H.
  1996, \apj, 460, 1034

\bibitem[{{Fletcher} \& {Hudson}(2008)}]{Fletcher2008}
{Fletcher}, L., \& {Hudson}, H.~S. 2008, \apj, 675, 1645

\bibitem[{{Fontenla} {et~al.}(1993){Fontenla}, {Avrett}, \&
  {Loeser}}]{Fontenla1993}
{Fontenla}, J.~M., {Avrett}, E.~H., \& {Loeser}, R. 1993, \apj, 406, 319

\bibitem[{{Golub} {et~al.}(2007){Golub}, {Deluca}, {Austin}, {Bookbinder},
  {Caldwell}, {Cheimets}, {Cirtain}, {Cosmo}, {Reid}, {Sette}, {Weber},
  {Sakao}, {Kano}, {Shibasaki}, {Hara}, {Tsuneta}, {Kumagai}, {Tamura},
  {Shimojo}, {McCracken}, {Carpenter}, {Haight}, {Siler}, {Wright}, {Tucker},
  {Rutledge}, {Barbera}, {Peres}, \& {Varisco}}]{Golub2007}
{Golub}, L., {Deluca}, E., {Austin}, G., {Bookbinder}, J., {Caldwell}, D.,
  {Cheimets}, P., {Cirtain}, J., {Cosmo}, M., {Reid}, P., {Sette}, A., {Weber},
  M., {Sakao}, T., {Kano}, R., {Shibasaki}, K., {Hara}, H., {Tsuneta}, S.,
  {Kumagai}, K., {Tamura}, T., {Shimojo}, M., {McCracken}, J., {Carpenter}, J.,
  {Haight}, H., {Siler}, R., {Wright}, E., {Tucker}, J., {Rutledge}, H.,
  {Barbera}, M., {Peres}, G., \& {Varisco}, S. 2007, \solphys, 243, 63

\bibitem[{{Harrison} {et~al.}(2005){Harrison}, {Christensen}, {Craig},
  {Hailey}, {Baumgartner}, {Chen}, {Chonko}, {Cook}, {Koglin}, {Madsen},
  {Pivavoroff}, {Boggs}, \& {Smith}}]{Harrison2005}
{Harrison}, F.~A., {Christensen}, F.~E., {Craig}, W., {Hailey}, C.,
  {Baumgartner}, W., {Chen}, C.~M.~H., {Chonko}, J., {Cook}, W.~R., {Koglin},
  J., {Madsen}, K.-K., {Pivavoroff}, M., {Boggs}, S., \& {Smith}, D. 2005,
  Experimental Astronomy, 20, 131

\bibitem[{{Holman} {et~al.}(2003){Holman}, {Sui}, {Schwartz}, \&
  {Emslie}}]{Holman2003}
{Holman}, G.~D., {Sui}, L., {Schwartz}, R.~A., \& {Emslie}, A.~G. 2003, \apjl,
  595, L97

\bibitem[{{Kane}(1972)}]{Kane1972}
{Kane}, S.~R. 1972, \solphys, 27, 174

\bibitem[{{Karlick{\'y}} \& {Kosugi}(2004)}]{Karlicky2004}
{Karlick{\'y}}, M., \& {Kosugi}, T. 2004, \aap, 419, 1159

\bibitem[{{Krucker} {et~al.}(2007{\natexlab{a}}){Krucker}, {Hannah}, \&
  {Lin}}]{Krucker2007a}
{Krucker}, S., {Hannah}, I.~G., \& {Lin}, R.~P. 2007{\natexlab{a}}, \apjl, 671,
  L193

\bibitem[{{Krucker} {et~al.}(2007{\natexlab{b}}){Krucker}, {Kontar}, {Christe},
  \& {Lin}}]{Krucker2007}
{Krucker}, S., {Kontar}, E.~P., {Christe}, S., \& {Lin}, R.~P.
  2007{\natexlab{b}}, \apjl, 663, L109

\bibitem[{{Krucker} \& {Lin}(2008)}]{Krucker2008}
{Krucker}, S., \& {Lin}, R.~P. 2008, \apj, 673, 1181

\bibitem[{{Kundu} {et~al.}(1995){Kundu}, {Raulin}, {Nitta}, {Hudson},
  {Shimojo}, {Shibata}, \& {Raoult}}]{Kundu1995}
{Kundu}, M.~R., {Raulin}, J.~P., {Nitta}, N., {Hudson}, H.~S., {Shimojo}, M.,
  {Shibata}, K., \& {Raoult}, A. 1995, \apjl, 447, L135+

\bibitem[{{Lin}(1985)}]{Lin1985}
{Lin}, R.~P. 1985, \solphys, 100, 537

\bibitem[{{Lin} {et~al.}(2002){Lin}, {Dennis}, {Hurford}, {Smith}, {Zehnder},
  {Harvey}, {Curtis}, {Pankow}, {Turin}, {Bester}, {Csillaghy}, {Lewis},
  {Madden}, {van Beek}, {Appleby}, {Raudorf}, {McTiernan}, {Ramaty}, {Schmahl},
  {Schwartz}, {Krucker}, {Abiad}, {Quinn}, {Berg}, {Hashii}, {Sterling},
  {Jackson}, {Pratt}, {Campbell}, {Malone}, {Landis}, {Barrington-Leigh},
  {Slassi-Sennou}, {Cork}, {Clark}, {Amato}, {Orwig}, {Boyle}, {Banks},
  {Shirey}, {Tolbert}, {Zarro}, {Snow}, {Thomsen}, {Henneck}, {McHedlishvili},
  {Ming}, {Fivian}, {Jordan}, {Wanner}, {Crubb}, {Preble}, {Matranga}, {Benz},
  {Hudson}, {Canfield}, {Holman}, {Crannell}, {Kosugi}, {Emslie}, {Vilmer},
  {Brown}, {Johns-Krull}, {Aschwanden}, {Metcalf}, \& {Conway}}]{Lin2002}
{Lin}, R.~P., {Dennis}, B.~R., {Hurford}, G.~J., {Smith}, D.~M., {Zehnder}, A.,
  {Harvey}, P.~R., {Curtis}, D.~W., {Pankow}, D., {Turin}, P., {Bester}, M.,
  {Csillaghy}, A., {Lewis}, M., {Madden}, N., {van Beek}, H.~F., {Appleby}, M.,
  {Raudorf}, T., {McTiernan}, J., {Ramaty}, R., {Schmahl}, E., {Schwartz}, R.,
  {Krucker}, S., {Abiad}, R., {Quinn}, T., {Berg}, P., {Hashii}, M.,
  {Sterling}, R., {Jackson}, R., {Pratt}, R., {Campbell}, R.~D., {Malone}, D.,
  {Landis}, D., {Barrington-Leigh}, C.~P., {Slassi-Sennou}, S., {Cork}, C.,
  {Clark}, D., {Amato}, D., {Orwig}, L., {Boyle}, R., {Banks}, I.~S., {Shirey},
  K., {Tolbert}, A.~K., {Zarro}, D., {Snow}, F., {Thomsen}, K., {Henneck}, R.,
  {McHedlishvili}, A., {Ming}, P., {Fivian}, M., {Jordan}, J., {Wanner}, R.,
  {Crubb}, J., {Preble}, J., {Matranga}, M., {Benz}, A., {Hudson}, H.,
  {Canfield}, R.~C., {Holman}, G.~D., {Crannell}, C., {Kosugi}, T., {Emslie},
  A.~G., {Vilmer}, N., {Brown}, J.~C., {Johns-Krull}, C., {Aschwanden}, M.,
  {Metcalf}, T., \& {Conway}, A. 2002, \solphys, 210, 3

\bibitem[{{Lin} \& {Hudson}(1971)}]{Lin1971}
{Lin}, R.~P., \& {Hudson}, H.~S. 1971, \solphys, 17, 412

\bibitem[{{Massone} {et~al.}(2004){Massone}, {Emslie}, {Kontar}, {Piana},
  {Prato}, \& {Brown}}]{Massone2004}
{Massone}, A.~M., {Emslie}, A.~G., {Kontar}, E.~P., {Piana}, M., {Prato}, M.,
  \& {Brown}, J.~C. 2004, \apj, 613, 1233

\bibitem[{{Nitta} {et~al.}(2008){Nitta}, {Mason}, {Wiedenbeck}, {Cohen},
  {Krucker}, {Hannah}, {Shimojo}, \& {Shibata}}]{Nitta2008}
{Nitta}, N.~V., {Mason}, G.~M., {Wiedenbeck}, M.~E., {Cohen}, C.~M.~S.,
  {Krucker}, S., {Hannah}, I.~G., {Shimojo}, M., \& {Shibata}, K. 2008, \apjl,
  675, L125

\bibitem[{{Paesold} {et~al.}(2001){Paesold}, {Benz}, {Klein}, \&
  {Vilmer}}]{Paesold2001}
{Paesold}, G., {Benz}, A.~O., {Klein}, K.-L., \& {Vilmer}, N. 2001, \aap, 371,
  333

\bibitem[{{Pick} {et~al.}(2006){Pick}, {Forbes}, {Mann}, {Cane}, {Chen},
  {Ciaravella}, {Cremades}, {Howard}, {Hudson}, {Klassen}, {Klein}, {Lee},
  {Linker}, {Maia}, {Mikic}, {Raymond}, {Reiner}, {Simnett}, {Srivastava},
  {Tripathi}, {Vainio}, {Vourlidas}, {Zhang}, {Zurbuchen}, {Sheeley}, \&
  {Marqu{\'e}}}]{Pick2006}
{Pick}, M., {Forbes}, T.~G., {Mann}, G., {Cane}, H.~V., {Chen}, J.,
  {Ciaravella}, A., {Cremades}, H., {Howard}, R.~A., {Hudson}, H.~S.,
  {Klassen}, A., {Klein}, K.~L., {Lee}, M.~A., {Linker}, J.~A., {Maia}, D.,
  {Mikic}, Z., {Raymond}, J.~C., {Reiner}, M.~J., {Simnett}, G.~M.,
  {Srivastava}, N., {Tripathi}, D., {Vainio}, R., {Vourlidas}, A., {Zhang}, J.,
  {Zurbuchen}, T.~H., {Sheeley}, N.~R., \& {Marqu{\'e}}, C. 2006, Space Science
  Reviews, 123, 341

\bibitem[{{Priest} \& {Forbes}(1986)}]{Priest2002}
{Priest}, E.~R., \& {Forbes}, T.~G. 1986, \jgr, 91, 5579

\bibitem[{{Ramsey} {et~al.}(2000){Ramsey}, {Alexander}, {Apple}, {Austin},
  {Benson}, {Dietz}, {Elsner}, {Engelhaupt}, {Kolodziejczak}, {O'Dell},
  {Speegle}, {Swartz}, {Weisskopf}, \& {Zirnstein}}]{Ramsey2000}
{Ramsey}, B.~D., {Alexander}, C.~D., {Apple}, J.~A., {Austin}, R.~A., {Benson},
  C.~M., {Dietz}, K.~L., {Elsner}, R.~F., {Engelhaupt}, D.~E., {Kolodziejczak},
  J.~J., {O'Dell}, S.~L., {Speegle}, C.~O., {Swartz}, D.~A., {Weisskopf},
  M.~C., \& {Zirnstein}, G. 2000, in Presented at the Society of Photo-Optical
  Instrumentation Engineers (SPIE) Conference, Vol. 4138, Proc. SPIE Vol. 4138,
  p. 147-153, X-Ray Optics, Instruments, and Missions IV, Richard B. Hoover;
  Arthur B. Walker; Eds., ed. R.~B. {Hoover} \& A.~B. {Walker}, 147--153

\bibitem[{{Raulin} {et~al.}(1996){Raulin}, {Kundu}, {Nitta}, \&
  {Raoult}}]{Raulin1996}
{Raulin}, J.~P., {Kundu}, M.~R., {Nitta}, N., \& {Raoult}, A. 1996, \apj, 472,
  874

\bibitem[{{Saint-Hilaire} \& {Benz}(2002)}]{PSH2002}
{Saint-Hilaire}, P., \& {Benz}, A.~O. 2002, \solphys, 210, 287

\bibitem[{{Schwartz} {et~al.}(2002){Schwartz}, {Csillaghy}, {Tolbert},
  {Hurford}, {Mc Tiernan}, \& {Zarro}}]{Schwartz2002}
{Schwartz}, R.~A., {Csillaghy}, A., {Tolbert}, A.~K., {Hurford}, G.~J., {Mc
  Tiernan}, J., \& {Zarro}, D. 2002, \solphys, 210, 165

\bibitem[{{Shibata} {et~al.}(1992){Shibata}, {Ishido}, {Acton}, {Strong},
  {Hirayama}, {Uchida}, {McAllister}, {Matsumoto}, {Tsuneta}, {Shimizu},
  {Hara}, {Sakurai}, {Ichimoto}, {Nishino}, \& {Ogawara}}]{Shibata1992}
{Shibata}, K., {Ishido}, Y., {Acton}, L.~W., {Strong}, K.~T., {Hirayama}, T.,
  {Uchida}, Y., {McAllister}, A.~H., {Matsumoto}, R., {Tsuneta}, S., {Shimizu},
  T., {Hara}, H., {Sakurai}, T., {Ichimoto}, K., {Nishino}, Y., \& {Ogawara},
  Y. 1992, \pasj, 44, L173

\bibitem[{{Spitzer}(1962)}]{Spitzer1962}
{Spitzer}, L. 1962, {Physics of Fully Ionized Gases} (Physics of Fully Ionized
  Gases, New York: Interscience (2nd edition), 1962)

\bibitem[{{Sprangle} \& {Vlahos}(1983)}]{Sprangle1983}
{Sprangle}, P., \& {Vlahos}, L. 1983, \apjl, 273, L95

\bibitem[{{Vr{\v s}nak} {et~al.}(2003){Vr{\v s}nak}, {Warmuth}, {Mari{\v
  c}i{\'c}}, {Otruba}, \& {Ru{\v z}djak}}]{Vrsnak2003}
{Vr{\v s}nak}, B., {Warmuth}, A., {Mari{\v c}i{\'c}}, D., {Otruba}, W., \&
  {Ru{\v z}djak}, V. 2003, \solphys, 217, 187

\bibitem[{{Wang} {et~al.}(2006){Wang}, {Lin}, {Krucker}, \&
  {Gosling}}]{Wang2006}
{Wang}, L., {Lin}, R.~P., {Krucker}, S., \& {Gosling}, J.~T. 2006, \grl, 33,
  3106

\bibitem[{{Wentzel}(1982)}]{Wentzel1982}
{Wentzel}, D.~G. 1982, \apj, 256, 271

\end{thebibliography}

\clearpage

\begin{table*}[ht!]
	\caption{GOES 10 response for LO/HI channels, in W/m$^2$. The beams have $\delta$=4, $E_1$=10 keV.}
	\centering
	\begin{tabular}{ccc}
	\tableline\tableline
		$F_1$				& ``down'' beam						& ``up'' beam	\\
	\tableline
		2.7$\times$10$^{36}$ e$^-$/s	&	1.0$\times$10$^{-7}$ (``B1.0'')/ 3.7$\times$10$^{-8}$		& 4.4$\times$10$^{-8}$ (``A4.4'')/ 1.5$\times$10$^{-8}$		\\
		1.0$\times$10$^{34}$ e$^-$/s	&	3.8$\times$10$^{-10}$  (``A0.04'')/ 1.4$\times$10$^{-10}$	& 1.6$\times$10$^{-10}$ (``A0.02'')/ 5.7$\times$10$^{-11}$		\\
	\tableline
	\end{tabular}
	\label{tab:goesresp}
	\end{table*}

	\begin{table*}[ht!]
	\caption{Temperatures and emission measures derived from fluxes computed in Table \ref{tab:goesresp}:}
	\centering
	\begin{tabular}{ccc}
	\tableline\tableline
		$F_1$				& ``down'' beam						& ``up'' beam	\\
	\tableline
		2.7$\times$10$^{36}$ e$^-$/s	&	Mewe: 23.8 MK, 6.6$\times$10$^{-46}$ cm$^{-3}$		& Mewe: 22.0 MK, 3.1$\times$10$^{-46}$ cm$^{-3}$ \\
						&	Chianti: 21.5 MK, 4.2$\times$10$^{-46}$ cm$^{-3}$	& Chianti: 20.2 MK, 1.9$\times$10$^{-46}$ cm$^{-3}$ \\
		\\
		1.0$\times$10$^{34}$ e$^-$/s	&	Mewe: 23.8 MK, 2.5$\times$10$^{-44}$ cm$^{-3}$		& Mewe: 23.0 MK, 1.1$\times$10$^{-44}$ cm$^{-3}$ \\
	\tableline
	\end{tabular}
	\label{tab:goesTEM}
	\end{table*}

\begin{table*}[ht!]
	\caption{Temperatures and emission measures derived from our coronal heating model (``down'' beam nonexistant).
		The first case (first line) leads to a beam density of the same order as the ambiant plasma, an unlikely situation.
		It has been kept for completeness' sake.}
	\centering
	\begin{tabular}{ccc}
	\tableline\tableline
		$F_1$				&  S			&	T [MK], EM [cm$^{-3}$]	\\
	\tableline
		2.7$\times$10$^{36}$ e$^-$/s	&  10$^{16}$ cm$^2$	&	89, 1.4$\times$10$^{44}$	\\
						&  10$^{18}$ cm$^2$	&	8.1, 1.4$\times$10$^{46}$\\
		1.0$\times$10$^{34}$ e$^-$/s	&  10$^{16}$ cm$^2$	&	4.5, 1.4$\times$10$^{44}$\\
						&  10$^{18}$ cm$^2$	&	2.4, 1.4$\times$10$^{46}$\\
	\tableline
	\end{tabular}
	\label{tab:beamheatingTEM}
	\end{table*}

	\begin{table*}[ht!]
	\caption{Required RHESSI Observing Summary countrates (counts/s/detector) above background to detect an ``up'' electron beam (in the absence of any other emission), and corresponding number of beam electrons.
		The typical background countrates for the 3--6 keV band and the 6--12 keV bands are $\approx$6 and  $\approx$15 counts/s/detector, respectively.}
	\centering
	\tiny
	\begin{tabular}{c|c|c|c|c|c|c|c|c|c|c}
	\tableline\tableline
		RHESSI  			&  detection		&	\multicolumn{3}{c|}{$\Delta t$=4 s}		& \multicolumn{3}{c|}{$\Delta t$=30 s}		&  \multicolumn{3}{c}{$\Delta t$=4 min}	\\
		energy band			&  level		&	cts/s/det & 10$^{34}$ e-/s &  10$^{35}$ e-	&cts/s/det & 10$^{34}$ e-/s &  10$^{35}$ e-	&	cts/s/det & 10$^{34}$ e-/s &  10$^{35}$ e-	\\
	\tableline
		3--6 keV			& 3-$\sigma$		&	4.7 & 5.5 & 2.2		&	1.7 & 1.9 & 5.8	&	0.6 & 0.68 & 16.3	\\	
						& 5-$\sigma$		&	13.3 & 15.5 & 6.2	&	4.6 & 5.4 & 16.2&	1.6 & 1.9 & 45.3	\\	
	\tableline
		6--12 keV			& 3-$\sigma$		&	7.3 & 1.5 & 0.6		&	2.6 & 0.5 & 1.5	&	0.9 & 0.18 & 4.3	\\	
						& 5-$\sigma$		&	20.5 & 4.0 & 1.6	&	7.3 & 1.4 & 4.3	&	2.6 & 0.5 & 12.0	\\	
	\tableline
	\end{tabular}
	\label{tab:lcdetection}
	\end{table*}

\clearpage

\end{document}